\documentclass[
prl, twocolumn,
superscriptaddress,
nofootinbib,
 amsmath,amssymb,
 aps,
floatfix
]{revtex4-1}

\usepackage{times}

\usepackage{amsmath}
\usepackage{amsfonts}
\usepackage{amssymb}
\usepackage{graphicx}

\usepackage{graphicx}
\usepackage{gensymb} 
\usepackage{dcolumn}
\usepackage{bm}
\usepackage{amsmath}
\usepackage{graphicx}
\usepackage{yfonts}
\usepackage{float}
\usepackage{mathtools}
\usepackage{comment}
\DeclareMathOperator{\tr}{Tr}

\usepackage{notes2bib}
\bibnotesetup{
note-name = ,
use-sort-key = false
}

\newcommand{\bra}[1]{\left\langle #1 \right|}
\newcommand{\ket}[1]{\left| #1 \right\rangle}

\newcommand{\ketbra}[2]{\left|#1\middle\rangle\middle\langle#2\right|}

\usepackage[normalem]{ulem}
\usepackage{color}

\newcommand\m[1]{\begin{pmatrix}#1\end{pmatrix}}

\begin{document} 
\preprint{APS/123-QED}

\title{Communicating via ignorance:  Increasing communication capacity via superposition of order}%

\author{K. Goswami} 
\email{k.goswami@uq.edu.au}
\affiliation{Centre for Engineered Quantum Systems, School of Mathematics and Physics, University of Queensland, QLD 4072 Australia}
\author{Y. Cao}
\affiliation{Centre for Quantum Dynamics, Griffith University, Queensland 4111, Australia}
\affiliation{State Key Laboratory of Networking and Switching Technology, Beijing University of Posts and Telecommunications,Beijing 100876, China}
\author{G. A. Paz-Silva}
\affiliation{Centre for Quantum Dynamics, Griffith University, Queensland 4111, Australia}
\author{J. Romero}
\email{m.romero@uq.edu.au}
\affiliation{Centre for Engineered Quantum Systems, School of Mathematics and Physics, University of Queensland, QLD 4072 Australia}
\author{A. G. White}
\affiliation{Centre for Engineered Quantum Systems, School of Mathematics and Physics, University of Queensland, QLD 4072 Australia}

\date{\today}

\date{\today}
\begin{abstract}
Classically, no information can be transmitted through a depolarising, that is a completely noisy, channel.  We show that by combining a depolarising channel with another channel in an indefinite causal order---that is, when there is superposition of the order that these two channels were applied---it becomes possible to transmit significant information. We consider two limiting cases. When both channels are fully-depolarising, the ideal limit is communication of 0.049 bits; experimentally we achieve $(3.4{\pm}0.2){\times}10^{-2}$ bits.  When one channel is fully-depolarising, and the other is a known unitary, the ideal limit is communication of 1 bit. We experimentally achieve 0.64${\pm}$0.02 bits. Our results offer intriguing possibilities for future communication strategies beyond conventional quantum Shannon theory.  
\end{abstract}
\maketitle

Noise is ubiquitous: communication protocols aim to optimise the amount of information that can be sent through a channel with a given amount of noise. In the limit of a  completely noisy channel, no information can be transmitted \cite{Shannon48}. This is true even with a single quantum channel \cite{Preskill2016}. Surprisingly, quantum physics offers strategies to transmit information in the scenario of two noisy channels, e.g.via superposition of path \cite{abbott2016, Gisin2005,Chiribella_path}, or via superposition of order \cite{Ebler2017,Salek2018,Chiribella2018}. In these strategies the superposition in a control qubit determines the superposition in path or causal order. Placing two completely noisy channels in a superposition of paths---that is in different arms of an interferometer---allows some classical information to be communicated, at least 0.16 bits when the paths are equally weighted. Placing them instead in a superposition of causal order---that is the order in which the channels are applied is indefinite---allows up to 0.049 bits to be communicated when the orders are equally   weighted \cite{Ebler2017}. Here we show that using superposition of causal order, a greater communication advantage can be achieved than superposition of paths \cite{abbott2016}, ideally up to 100$\%$ of information.


We use the quantum switch, a physically realisable process that simulates the superposition of causal order, which has been  implemented in several photonic experiments \cite{Procopio2014, rubino2017, Goswami18, Wei2018, guo2018experimental}. We follow the experimental setup of \cite{Goswami18}. 
We analyse, and experimentally demonstrate, communication through various combinations of noisy and unitary channels in an indefinite causal order. We provide an example where perfect transmission is possible, given the freedom to choose the measurement basis of the control that determines the order. We outline the mathematical description of the channels and provide an experimental method to estimate the information-theoretic advantage quantified by the Holevo capacity, $\chi$ \cite{Holevo73}.
\begin{figure}[!t]
\begin{center}
\includegraphics[width=\columnwidth]{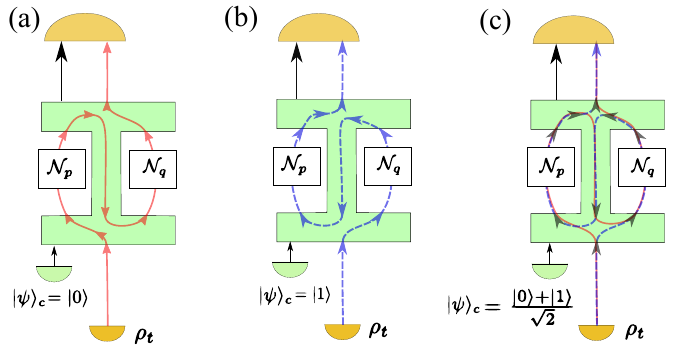}
\vspace{-2mm}
\caption{Classical communication through  indefinite causal order. The sender maps the classical message bit to a quantum state $\rho_t$---the target qubit. $\rho_t$ passes through two noisy channels $\mathcal{N}_p$ and $\mathcal{N}_q$. The order of the channels is controlled by a control qubit $\ket{\psi_c}$. (a) When the control qubit $\rho_c{=}\ket{\psi_{c}}\bra{\psi_{c}}$ is off, i.e. $\ket{\psi_{c}}{=}\ket{0}$, the channels have a definite order $\mathcal{N}_q {\circ} \mathcal{N}_p$. (b) When the control qubit is on, $\ket{\psi_{c}}{=}\ket{1}$, the order is $\mathcal{N}_{p} {\circ} \mathcal{N}_{q}$.   (c) When the control is in a superposition, $\ket{\psi_{c}}{=}(\ket{0}{+}\ket{1})/\sqrt{2}$, the channels have an indefinite order.  \vspace{0 mm}}
\label{fig:Qswitch}
\end{center}
\end{figure}

We depict the communication protocol in the Fig. \ref{fig:Qswitch}. The sender maps a classical message into a quantum state $\rho_t{=} \ketbra{\psi_t}{\psi_t}$, which we will refer to as the target qubit. This state passes through two generalised Pauli channels $\mathcal{N}_p$ and $\mathcal{N}_q$. We describe the noisy channels acting on the target qubit $\rho_t$ as
\begin{align}
\mathcal{N}_p(\rho_t){=}\sum_{i=0}^3p_{i}\sigma_{i}\rho_t\sigma_{i}^\dagger \\ 
\mathcal{N}_q(\rho_t){=}\sum_{i=0}^3q_{i}\sigma_{i}\rho_t\sigma_{i}^\dagger,
 \label{channel}
 \end{align}

 \noindent where $\sum_i p_{i} {=} \sum_i q_{i} {=} 1$. As the equation suggests, each Pauli channel is a probabilistic mixture of all three Pauli errors via the identifications: $\sigma_{1} {\equiv} \sigma_x$ (bit flip); $\sigma_{3} {\equiv} \sigma_z$ (phase flip); $\sigma_{2} {\equiv} \sigma_y$ (combination of bit flip and phase flip); and $\sigma_{0} {\equiv} I$.  The order that these two channels are applied to $\rho_t$ is selected by a control qubit $\rho_c$. If the control is off, $\ket{0}_c$, then the order is $\mathcal{N}_{q} {\circ} \mathcal{N}_{p}$, i.e. $\mathcal{N}_{p}$ is before $\mathcal{N}_{q}$. If the control is on, $\ket{1}_c$, then $\mathcal{N}_{p} {\circ} \mathcal{N}_{q}$. However, if the control qubit is in a superposition state,  $\ket{\psi}_{c} {=} (\ket{0} {+} \ket{1})_c/\sqrt{2}$, the channels $\mathcal{N}_{p}$ and $\mathcal{N}_{q}$ have indefinite causal order. Specifically, given a target qubit $\rho_t$ and a control qubit initially in the state $\rho_{c}{=}|\psi_{c}\rangle\langle\psi_{c}|$, where $\ket{\psi_{c}}{=}\sqrt{\gamma}|0 \rangle {+}\sqrt{1{-}\gamma}|1 \rangle $,  the total output state of the switch becomes
 
 \begin{align}
 \textswab{s:}[\mathcal{N}_{p},\mathcal{N}_{q}](\rho_c{\otimes}\rho_t){=}\sum_{i,j}K_{ij} (\rho_c{\otimes}\rho_t)K_{ij}^{\dagger}  
 \end{align}
 with
\begin{align}
 K_{ij}{=}p_iq_j(\ketbra{0}{0} \otimes \sigma_i\sigma_j + \ketbra{1}{1} \otimes \sigma_j\sigma_i)
 \end{align}
 In matrix representation, the overall output state becomes
\begin{equation}
\textswab{s:}[\mathcal{N}_{p},\mathcal{N}_{q}](\rho_c{\otimes}\rho_t){=}\left(\begin{array}{cc}
A&B\\
B&\tilde{A}\\
\end{array} \right),
\label{output_AB_Matrix}
\end{equation}
with, 
\begin{equation}
\begin{array}{l}
A{=}\gamma \mathcal{N}_q\circ \mathcal{N}_p(\rho_t),\\
\tilde{A}{=}(1-\gamma)\mathcal{N}_p\circ \mathcal{N}_q(\rho_t),\\
B{=}\sqrt{\gamma(1-\gamma)}(\epsilon_+(\rho_t)-\epsilon_-(\rho_t)),\\
  \end{array}
\label{A_and_B}
\end{equation}
where ($\epsilon_-$) $\epsilon_+$ represents an auxiliary trace non-preserving map composed of all the operators from $\mathcal{N}_p$ and $\mathcal{N}_q$ that (anti-commute) commute. That is, 
\begin{align}
 \epsilon_+(\rho_t) &{=}\sum_{i=0}^3p_iq_i\, \rho_t + \sum_{i=0}^3 r_{0i}\,\sigma_i\rho_t\sigma_i^\dagger \\
 \epsilon_-(\rho_t) & {=}r_{12}\,{\sigma_3}\rho_t\sigma_3^\dagger {+}r_{23}\,\sigma_1\rho_t\sigma_1^\dagger {+} r_{31}\,\sigma_2\rho_t\sigma_2^\dagger ,
\label{commutator_anticommutator_channel}
\end{align}
with $r_{ij} {=} p_iq_j{+}p_jq_i$. Note that $\mathcal{N}_p{\circ} \mathcal{N}_q(\rho_t) {=} \epsilon_+(\rho_t)+\epsilon_-(\rho_t)  {=} \mathcal{N}_q{\circ }\mathcal{N}_p(\rho_t)$, any definite order of $\mathcal{N}_p$ and $\mathcal{N}_q$ will have the same effect on the target qubit. To take an extreme example, if $\mathcal{N}_p$ and $\mathcal{N}_q$ were both completely depolarising, either of the definite orders $\mathcal{N}_p{\circ} \mathcal{N}_q(\rho_t)$ or $\mathcal{N}_q{\circ} \mathcal{N}_p(\rho_t)$  will completely scramble the target qubit. 


Interestingly, the output of a quantum switch (Eqs. \ref{output_AB_Matrix}-\ref{A_and_B}) implies that some information is contained in the control qubit. Depending on the outcome of a measurement in the $\sigma_1^c$ basis of the control qubit,  we obtain either of the conditional states $\epsilon_{+}(\rho_t)$ or $\epsilon_{-}(\rho_t)$. This means that a $\sigma_1^c$ measurement allows us to estimate information encoded in the target qubit, and confirm whether there is a communication advantage when $\mathcal{N}_p$ and $\mathcal{N}_q$ are in an indefinite order.

In our quantum switch experiment,  the control qubit is the polarisation of light. A $\sigma_1^c$ measurement in this case is equivalent to a Stokes measurement $S_2(\textswab{s:}[\mathcal{N}_{p},\mathcal{N}_{q}])$ at the output of the quantum switch $\textswab{s:}[\mathcal{N}_{p},\mathcal{N}_{q}]$. As Eq. \ref{channel} suggests, the channels $\mathcal{N}_p$ and $\mathcal{N}_q$ can be constructed from combinations of Pauli operations $\{\sigma_i \}$ acting on the target qubit. In the supplementary material, we show how  $S_2(\textswab{s:}[\mathcal{N}_{p},\mathcal{N}_{q}])$ can be calculated from Stokes measurements $S_2(\textswab{s:}[\sigma_i,\sigma_j])$ at the output of a quantum switch of $\sigma_i$ and $\sigma_j$, $\textswab{s:}[\sigma_i,\sigma_j]$.  This means that rather than physically implementing  $\textswab{s:}[\mathcal{N}_{p},\mathcal{N}_{q}]$, we can simply use Stokes measurements from $\textswab{s:}[\sigma_i,\sigma_j]$, so long as we keep track of the Pauli operations we perform, in effect using an additional memory.  The detailed description of our quantum switch and how we implemented the operations $\{\sigma_i \}$ on the transverse spatial mode---our target qubit---can be found in the Supplementary Material.  

The Stokes parameters $S_2(\textswab{s:}[\sigma_i,\sigma_j])$ are important because we use these to obtain the output control qubit $\tilde{\rho}_c$ after tracing out the target qubit. With knowledge of $\tilde{\rho}_c$, we can then calculate the Holevo capacity $\chi(\textswab{s:}[\mathcal{N}_{p},\mathcal{N}_{q}])$---a measure of the maximum amount of classical information that can be transferred through our arrangement of indefinitely ordered channels.  This is  given by ~\cite{Ebler2017}: 
\begin{equation}
\chi(\textswab{s:}[\mathcal{N}_{p},\mathcal{N}_{q}]) {=} 1 {+} H(\tilde{\rho}_c)-H^{\text{min}}(\textswab{s:}[\mathcal{N}_{p},\mathcal{N}_{q}]), 
\label{Holevo}
\end{equation}
\noindent where $H^{\text{min}}(\textswab{s:}[\mathcal{N}_{p},\mathcal{N}_{q}])$ is the minimum entropy of the two-qubit output of the quantum switch, and $H(\tilde{\rho}_c)$ is the von-Neumann entropy of the output control qubit $\tilde{\rho_c}$ (see Supplementary Material for explicit expressions in terms of $S_2(\textswab{s:}[\sigma_i,\sigma_j])$ ). 

\begin{table}[!t]
\begin{center}
\vspace{2mm}

    \begin{tabular}{| c | c | c | c | c | c |}
    \hline
     $\ \sigma_i \ $ & $\ \sigma_j \ $ & $\,\tilde{\rho}_t $ & $\, \tilde{\rho}_c $ & $\,S_2(\textswab{s:}[\sigma_i,\sigma_j])^{\textup{theor.}}\,$ & $\, S_2(\textswab{s:}[\sigma_i,\sigma_j])^{\textup{exp.}}\,$ \\ \hline
   
    ${\sigma}_0$ & ${\sigma}_0$  &    $\phantom{-i}\ket{1}$            & D &\phantom{-}1 & $\,$\phantom{-} 0.8547 $\pm\, 0.0006$ $\,$ \\ 
        & ${\sigma}_1$  &     $\phantom{-i}\ket{0}$         & D &\phantom{-}1 &\phantom{-}0.8718 $\pm\, 0.0005$\\ 
        & ${\sigma}_2$   &       $-i\ket{0}$      & D &\phantom{-}1 &\phantom{-}0.8792 $\pm\, 0.0005$\\ 
        & ${\sigma}_3$    &    \phantom{i}$-\ket{1}$           & D& \phantom{-}1&\phantom{-}0.8823 $\pm\, 0.0005$\\ 
      ${\sigma}_1$ & ${\sigma}_0$  &     $\phantom{-i}\ket{0}$                 & D & \phantom{-}1 & \phantom{-}0.8459 $\pm\, 0.0006$ \\ 
       & ${\sigma}_1$  &       $\phantom{-i}\ket{1}$       & D &\phantom{-}1 &\phantom{-}0.8439 $\pm\, 0.0007$\\ 
       & ${\sigma}_2$   &       $-i\ket{1}$          & A &-1 &-0.8434 $\pm\, 0.0006$\\ 
       & ${\sigma}_3$    &             \phantom{i}$-\ket{0}$   & A &-1 &-0.8540 $\pm\, 0.0007$\\ 
          
     ${\sigma}_2$  & ${\sigma}_0$ &      $-i\ket{0}$ &D &\phantom{-}1 &\phantom{-}0.8473 $\pm\, 0.0007$\\ 
       & ${\sigma}_1$  &    $\phantom{-}i\ket{1}$          & A &-1 &-0.8600 $\pm\, 0.0005$\\ 
       & ${\sigma}_2$   &           $\phantom{-i}\ket{1}$     & D&\phantom{-}1 &\phantom{-}0.8447 $\pm\, 0.0006$ \\ 
       & ${\sigma}_3$    &        $\phantom{-}i\ket{0}$   & A & -1 &-0.8278 $\pm\, 0.0008$\\ 
       
      ${\sigma}_3$ & ${\sigma}_0$ &     \phantom{i}$-\ket{1}$           & D&\phantom{-}1 & \phantom{-}0.8316 $\pm\, 0.0006$\\ 
       & ${\sigma}_1$ &     $\phantom{-i}\ket{0}$          & A &-1 &-0.8228 $\pm\, 0.0006$\\ 
       & ${\sigma}_2$  &          $-i\ket{0}$    & A &-1 &-0.8575 $\pm\, 0.0006$\\ 
       & ${\sigma}_3$   &     $\phantom{-i}\ket{1}$       &  D &\phantom{-}1 &\phantom{-}0.8780 $\pm\, 0.0005$\\ 
    \hline
    \end{tabular}

    \caption{Data used to calculate the Holevo capacity using Eq. \ref{Holevo} for the different combinations of unitary operations $\sigma_i$ and $\sigma_j$. $\tilde{\rho}_t$ and $\tilde{\rho}_c$ are the output target and control state respectively. $S_2(\textswab{s:}[\sigma_i,\sigma_j])$ is the Stokes parameter obtained by measuring polarisation of the control qubit in the diagonal/anti-diagonal basis after the unitaries $\sigma_i$ and $\sigma_j$.
We list both the theoretically expected values, $S_2(\textswab{s:}[\sigma_i,\sigma_j])^{\textup{theor.}}$, and the experimentally measured values, $\, S_2(\textswab{s:}[\sigma_i,\sigma_j])^{\textup{exp.}}\,$, of this Stokes parameter. Error bars ($1\sigma$) were calculated by propagation of error on the individual Stokes parameter with Poissonian counting statistics. \vspace{-7mm}}
\label{table:Stokes_table}
\end{center}
\end{table}

\begin{figure}[h]
\begin{center}
\includegraphics[width=0.8 \columnwidth]{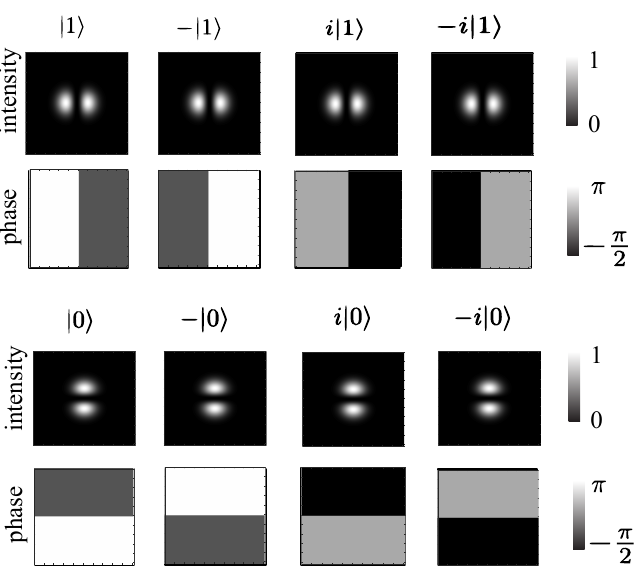}
\vspace{-3mm}
\caption{Predicted spatial mode of target qubit outputs, $\tilde{\rho}_t$, after the Pauli operations $\{\sigma_i,\sigma_j\}$ (see columns 1--3 of Table \ref{table:Stokes_table}). The input target qubit $\rho_t$ is $\ket{1} {=} \mathrm{HG}_{10}$, a first-order Hermite-Gaussian mode. \vspace{-5mm}
}
\label{fig:spatial_mode_and_target_qubit}
\end{center}
\end{figure}

In our experiment, we implemented the quantum switch with all 16 pairs of $\sigma_i$ and $\sigma_j$, as shown in Table \ref{table:Stokes_table}. We measured $S_2(\textswab{s:}[\sigma_i,\sigma_j])$, i.e. the diagonal and anti-diagonal components of the polarisation of the output light, for each pair $\sigma_i,\sigma_j$,  with the control qubit initialised to $\ket{\psi}_{c} {=} (\ket{0} {+} \ket{1})_c/\sqrt{2}$--- diagonally polarised light. We measure over 10 s; the measured rate at the control output is around 100,000 counts per second. Table \ref{table:Stokes_table} summarises the results. The first two columns are the ideal settings for the target operations on the transverse spatial mode, $\sigma_i,\sigma_j$, the third and fourth columns are the ideal output transformations, $\tilde{\rho}_t , \tilde{\rho}_c$ of the target and control qubits, respectively. Fig. \ref{fig:spatial_mode_and_target_qubit} shows the action of the transformations from column 3 on our choice of input target qubit. The last two columns are the ideal and measured values of the Stokes parameter of the control output. 
We minimise the uncertainty in each measurement by accumulating a large number of counts. Computer-controlled waveplates, with angular orientation uncertainty of $\pm (2.5 {\times} 10^{-4})^{\circ}$, are used to measure in the diagonal/anti-diagonal basis. Our measurements are limited by the non-ideal interferometric visibility in our switch. We calculate the average visibility from the values mentioned in Table \ref{table:Stokes_table}, which is $85{\pm} 2$\%. This is due to several factors. We use inverting prisms that have relatively coarse rotation accuracy $\pm 1^{\circ}$ to perform the unitary operations on the spatial mode. Moreover, each optical element is not perfectly flat, introducing wavefront distortions that limit the final visibility. 

\begin{figure}[!t]
\begin{center}
\includegraphics[width=\columnwidth]{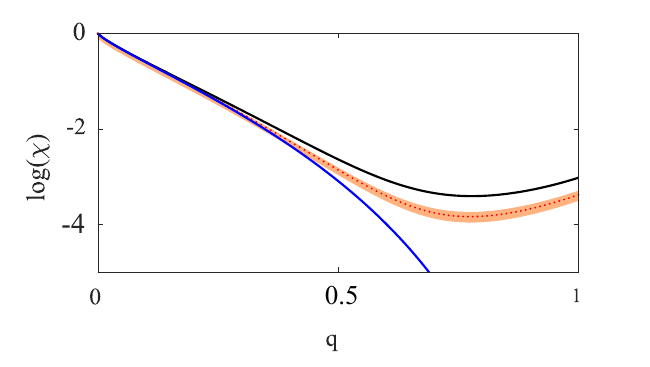}
\caption{ Logarithm of Holevo capacity $\chi$, of two identical depolarising channels, versus depolarising channel strength, $q$. The blue line is the predicted Holevo capacity for the case of definite order. The black line is the predicted Holevo capacity for the case of indefinite order. The red dots are the measured values for Holevo capacity in our quantum switch; in excellent agreement with the predicted Holevo capacity for our experimental visibilities $85 {\pm} 2$\%, shown by the orange shaded area. Note that the minimum measured Holevo capacity of $\chi{=} (2.1{\pm}0.2){\times}10^{-2}$ bits occurs at a depolarisation $q {=}0.78$, and that higher capacity of $\chi{=} (3.4{\pm}0.2){\times}10^{-2}$ bits occurs at full depolarisation, $q {=} 1$. \vspace{-6mm}
}
\label{depolarisation_graph}
\end{center}
\end{figure}

Now that we can calculate the output control qubit from Table  \ref{table:Stokes_table}, we can use Eq. \ref{Holevo} to estimate the Holevo capacity $\chi$.  We compare the experimental and predicted Holevo capacities for several combinations of channels. We first consider a quantum switch $\textswab{s:}[\mathcal{N},\mathcal{N}]$  of two depolarizing channels $\mathcal{N}$ of identical strengths $q$, i.e. in Eq.\ref{channel} we set all coefficients to $q/4$:
\begin{equation}
\begin{aligned}
\mathcal{N}&{=}(1{-}3q/4)\rho_t{+}q{/}4(\sigma_1\rho_t\sigma_1{+}\sigma_2\rho_t\sigma_2{+}\sigma_3\rho_t\sigma_3),  \\
\end{aligned}
\label{depolarising channels}
\end{equation}
\noindent This scenario was theoretically studied in Ref. \cite{Ebler2017}. Experimental and predicted Holevo capacities are shown in Fig. \ref{depolarisation_graph}, which plots the logarithm of $\chi$ against increasing $q$. The red circles are our measured values; the orange shaded region is the predicted Holevo capacity for visibilities of $85 {\pm} 2$\%; the black curve is the Holevo capacity for an ideal quantum switch. The blue curve is the ideal Holevo capacity for two depolarising channels in some definite order which---as expected--- decreases monotonically with increasing depolarising strength. In the limit of two fully depolarising channels, $q{=}1$, $\chi{=}0$ bits are transmitted using definite order. In this limit there is a clear advantage in using quantum channels: ideally $4.9{\times}10^{-2}$ bits can be transmitted, we measure $\chi{=}(3.4{\pm}0.2){\times}10^{-2}$ bits.  
This is a rather counterintuitive result as, individually none of the channels can transmit any information. 
  
This nonzero Holevo capacity can be understood intuitively from the fact that the output of a quantum switch with two depolarising channels is a statistical mixture of the output of a quantum switch with different Pauli operations $\{\sigma_i\}$. Some of these Pauli operations anti-commute, hence superpositions of the order of anti-commuting Pauli operations can preserve a finite amount of information in the target qubit.

This intuition helps us understand another striking prediction, which is that above some non-zero depolarisation strength the Holevo capacity will increase. In the ideal case, we see that Holevo capacity attains a minimum value of $3.3{\times}10^{-2}$ bits at $q{=} 0.78$ and then the capacity increases to the limit of $4.9{\times}10^{-2}$ bits at $q {=} 1$, in stark contrast to the classical case of definite causal order which decreases monotonically to zero. Experimentally, we see this increase in information capacity from $\chi{=}(2.1{\pm}0.2){\times}10^{-2}$ bits at $q{=}0.78$, i.e. at its worst absolute performance it is 13.5 $\sigma$ above the classical performance at that value of $q$. 

We look at Eq. \ref{Holevo} to understand this behaviour:  for low depolarisation strengths, the minimum entropy $H^{\text{min}}(\textswab{s:}[\mathcal{N}_{p},\mathcal{N}_{q}])$ increases more rapidly than the von Neumann entropy $H(\tilde{\rho}_c)$. This means that for low $q$, the rate of depolarisation of the composite system---target and control---is faster than the rate of depolarisation of the control qubit. However, at $q {\approx} 0.78$, $H(\tilde{\rho}_c)$ begins to increase more rapidly than $H^{\text{min}}(\textswab{s:}[\mathcal{N}_{p},\mathcal{N}_{q}])$ so the depolarisation rate of the control overtakes the depolarisation rate of the composite system, and the information revival occurs.

\begin{figure}[!h]
\begin{center}
\includegraphics[width=\columnwidth]{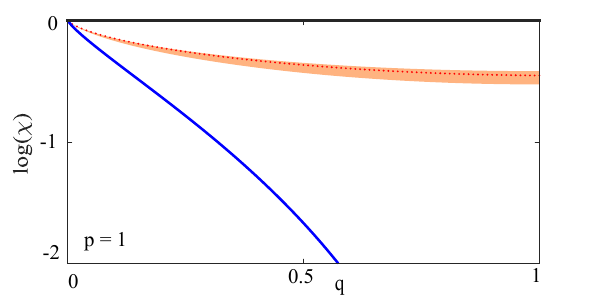}
\caption{Logarithm of Holevo capacity $\chi$, where one channel is a depolarising channel of varying strength $q$, and the other is a $\sigma_3$ channel, the channel in Eq. \eqref{extreme_case_channels} with $p{=}1$. The solid black line $(\mathrm{log}(\chi){=}0)$ is the theoretical predictions for Holevo capacity for indefinite causal order. The orange shade show the uncertainty due to non-ideal visibility in our experiment. The red dots are the experimentally measured data points and the blue line is the theoretical predictions for Holevo capacity for definite order. We show the special case for $p {=}1$, i.e. the dephasing channel becomes a $\sigma_3$ channel. Note that in this situation, indefinite causal order allows us to get full 1 bit information. Experimentally we measure a Holevo capacity of $0.64{\pm}0.02$ bits at $q{=}1$.  }
\vspace{-3mm}
\label{1bit_graph}
\end{center}
\end{figure}

Consequently, as a second example we consider the quantum switch $\textswab{s:}[\mathcal{N},\mathcal{M}]$ composed of a depolarizing channel $\mathcal{N}$ of strength $q$ (Eq. \ref{depolarising channels}) and a dephasing channel $\mathcal{M}$ of strength $p$ given by: 
\begin{equation}
\begin{aligned}
\mathcal{M}&{=}[(1-p)\rho_t{+}{p}\sigma_3\rho_t\sigma_3].
\end{aligned}
\label{extreme_case_channels}
\end{equation}
 When $q{=}1$, any definite order of these two channels results in a fully depolarizing channel, regardless of the value of $p$, thus $\chi{=}0$.  However, with the additional freedom to prepare and measure the control qubit, the Holevo capacity of these two channels in indefinite order, $\chi(\textswab{s:}[\mathcal{N},\mathcal{M}]){\equiv}\chi_{i}$ can be non-zero and is a function of $p$. In fact, this combination achieves the maximum value of 1 bit when $p{=}1$ regardless of the depolarisation strength $q$, as shown in Fig.~\ref{1bit_graph} (black line). We emphasise the contrast with the zero capacity of the definitely ordered channels at $q{=}1$, Fig.~\ref{1bit_graph} (blue line). 
 \\
 This unit Holevo capacity can be understood by noting that when $p{=}1$, $\mathcal{M}$ is a unitary channel and the conditional states become:
\begin{equation}
  \begin{array}{l}
\epsilon_+(\rho_t){=}(q/4)\rho_t+(1-3q/4) \sigma_3\rho_t\sigma_3,\\
\epsilon_-(\rho_t){=}q/4(\sigma_1\rho_t\sigma_1+\sigma_2\rho_t\sigma_2).
\end{array}
\end{equation}
and thus from Eq. \ref{Holevo} $\chi_{i} {=}1{+}H(\tilde{\rho_c}){-}H^{\text{min}}_{\rho_t} [\epsilon_{+}(\rho_t){\oplus}\epsilon_{-}(\rho_t)]$, where $H(\cdot)$ is again the von Neumann entropy and the minimisation is done over all possible target states. Notice that the states that minimize the entropy of $\epsilon_{+}(\rho_t){\oplus}\epsilon_{-}(\rho_t)$ are the eigenvectors of $\sigma_3$. Take $\rho{=}|0\rangle\langle0|$ as an example, $\epsilon_+(\rho_t){=}(1{-}q{/}2)|0\rangle\langle0|$, while $\epsilon_-(\rho_t)$, will be $(q/2)\ketbra{1}{1}$ in which case, $H[\epsilon_{+}(\rho){\oplus}\epsilon_{-}(\rho)]$ ${=}H(q{/}2)$, and thus the Holevo-capacity of the quantum switch becomes 1 bit \bibnote{This maximum advantage over definite order is also attained by having an indefinite order of one fully depolarising channel and a unitary $\sigma_2$ or $\sigma_3$ channel, when the input states are the eigenstates of $\sigma_2$ and $\sigma_3$, respectively.}. We compare our experimental and predicted $\chi_i$ in Fig. ~\ref{1bit_graph}, red dots are experimental values and the orange shaded regions accounts for non-ideal visibility. For full depolarisation at $q{=}1$,  we show $0.64{\pm}0.02$ bits compared to the ideal case of $\chi_i{=}1$.  Interestingly, this ideal capacity is strictly larger than the one achieved in Ref.\cite{Branciard2018} given the same channels in a path superposition. More strongly, in the Supplementary Material, we prove that these channels in a superposition of path cannot lead to unit capacity.  We also compare predicted and experimental $\chi_i$ for other dephasing strengths $p$ in the Supplementary Material. 

We note that when \cite{Ebler2017} proposed communication advantage from indefinite causal order they did not claim that the advantage is unique to coherent superposition of order. Subsequently, there has been an active discussion on the origin of the communication advantage \cite{Branciard2018, Guerin_2019,Hler_2019} with consensus yet to be reached.  The experimental developments in the present work and \cite{guo2018experimental}  are further motivatation to develop a well-defined resource communication theory featuring coherence.


Quantum mechanics allows indefinite causal order, which allows us to communicate up to 1 (0.049) bits of information through one (two) fully depolarising channels. This opens up interesting applications. 
Consider a situation where we wish to communicate through a field channel which is noisy in the target degree of freedom but unitary in the control degree of freedom, e.g. the atmosphere is noisy for spatial mode, but preserves polarisation. We can establish a lab channel that is unitary in both the control and target degrees of freedom, and place it in indefinite causal order with the field channel. The control system can be low-dimension---qubits---whereas the target degree of freedom may may be of higher dimension---qudits---so this enables communication through a channel that is otherwise completely noisy for qudits. Note that the field channel must be one that is naturally a return channel---i.e. the field channel exits and reenters the lab---since it is being used in indefinite causal order. The architecture presented in this paper can be useful for imaging through turbid media where the information being communicated is the modulation of the spatial mode. Possible examples are imaging through skin or establishing a return channel to a satellite, in both cases, the midpoint of the noisy return channel is providing modulation information that would normally be entirely lost.


We can also use this idea for secret sharing, where a specific combination of channels, when connected indefinitely, can transfer information between two parties, but in a scenario of eavesdropping, any intervention will break the `indefiniteness', and the message will remain scrambled.

\noindent \emph{Acknowledgements}. We thank Prof Howard Wiseman for suggesting this collaboration. We also thank Prof. Giulio Chiribella for a helpful discussion. KG thanks the organisers of the Quantum Information Structure Of Spacetime, 2020 workshop where valuable insights related to this work were discussed.  This work has been supported by: the Australian Research Council (ARC) by Centre of Excellence for Engineered Quantum Systems (EQUS, CE170100009), a DECRA grant (DE160100409), Advance Queensland Funding, and L'Oreal-UNESCO FWIS grant for JR; a DECRA grant (DE170100088) for GPS; YC is supported by a CSC scholarship; the University of Queensland by a Vice-Chancellor's Senior Research and Teaching Fellowship for AGW; and the John Templeton Foundation (the opinions expressed in this publication are those of the authors and do not necessarily reflect the views of the John Templeton Foundation).  We acknowledge the traditional owners of the land on which the University of Queensland is situated, the Turrbal and Jagera people.  

\bibliography{Depolbiblio}

\begin{thebibliography}{21}
\bibitem [{1()}]{1}%
  \BibitemOpen
  \href@noop {} {}\bibinfo {note} {This maximum advantage over definite order
  is also attained by having an indefinite order of one fully depolarising
  channel and a unitary $\sigma _2$ or $\sigma _3$ channel, when the input
  states are the eigenstates of $\sigma _2$ and $\sigma _3$,
  respectively.}\BibitemShut {Stop}%
\end{thebibliography}

\pagebreak
\renewcommand{\theequation}{S\arabic{equation}}
\setcounter{equation}{0}

\renewcommand{\thetable}{S\arabic{table}}
\setcounter{table}{0}

\renewcommand{\thefigure}{S\arabic{figure}}
\setcounter{figure}{0}

\newpage
\section*{Supplementary Material}
\subsection*{Estimating the Stokes parameter of the quantum switch}

\noindent With the Pauli decomposition discussed above,  we now show how measurements of the control qubit,  $\tilde{\rho}_c$,  can be used to estimate the Holevo capacity.
Since our control qubit is polarisation, we express the output control in terms of the Stokes vector $(S_1, S_2, S_3)$ \cite{Stokes1852}:
\begin{equation}
\tilde{\rho}_c {=} \frac{1}{2}\m{1 {+} S_1 & S_2 {+} i\,S_3\\ S_2{-}i\, S_3 & 1 -S_1}.
\label{stokes_definition}
\end{equation}

From Eq.\ref{output_AB_Matrix}, if we measure the output control state, $\tilde{\rho}_c$ becomes,
 
\begin{align}
\tilde{\rho}_c &{=} \m{\tr{(A)} & \tr{(B)}\\ \tr{(B)}\, & \tr{(\tilde{A}})} \nonumber \\
&{=}\m{\gamma & \tr{(B)}\\ \tr{(B)}\, & 1-\gamma},
\label{output_control_AB}
\end{align}
where the second equality comes from the fact that, Pauli channels are trace preserving. It is easy to see that $\tr{(B)}$ is a real number, so comparing Eq. \ref{stokes_definition} and \ref{output_control_AB} at $\gamma {=} 1/2$, we can see that $S_2 {=} 2\tr{(B)}$ and $S_1 {=} S_3 {=} 0$. Thus we  consider the effect of the quantum switch on the $S_2$ component of the control,  note that to distinguish the operations on the control and target, we use $\sigma_0^c$, $\sigma_1^c$, $\sigma_2^c$ and $\sigma_3^c$ for the control qubit and $\sigma_0$, $\sigma_1$, $\sigma_2$ and $\sigma_3$ for the target qubit,

\begin{equation}
\begin{aligned}
S_{2}(\textswab{s:}[\mathcal{N}_{p},\mathcal{N}_{q}]) & {=} \tr{\{(\sigma_1^c {\otimes} \sigma_0) \:. \textswab{s:}[\mathcal{N}_{p},\mathcal{N}_{q}]\} }\\
& {=} \sum_{i,j}\,p_iq_j\:\tr{\{(\sigma_1^c\otimes \sigma_0)\:.\textswab{s:}[\sigma_i,\sigma_j]\}}\\
& {=} \sum_{i,j}\,p_iq_j\:S_2(\textswab{s:}[\sigma_i,\sigma_j])
\label{Stokes}
\end{aligned}
\end{equation}
The above equation shows that with knowledge of the control qubit for individual combinations of $\sigma_i,\sigma_j$ we can get the $S_2$ of the switch with channels $\mathcal{N}_{p}$ and $\mathcal{N}_{q}$. 

\noindent From Eq. \ref{Stokes}, we get $\tilde{\rho}_c$, and we can calculate the von Neumann entropy $H(\tilde{\rho}_c)$, which is necessary to obtain the Holevo capacity as shown in Eq. \ref{Holevo}. The other quantity needed to evaluate \ref{Holevo} is the minimum entropy of the total output $H^{\text{min}}(\textswab{s:}[\mathcal{N}_{p},\mathcal{N}_{q}])$. In order to calculate this, let us first write the action of the quantum switch, in terms of individual combinations of Pauli operations:

\begin{eqnarray}
\textswab{s:}[\mathcal{N}_{p},\mathcal{N}_{q}] \ (\rho_c\otimes \rho_t) & {=} & \sum_{i,j}p_iq_j\: (\gamma \,\ketbra{0}{0}_c\,\otimes\,\sigma_i\sigma_j\rho_t\sigma_j^\dagger\sigma_i^\dagger \nonumber \\
& & \quad + (1{-}\gamma)\,\ketbra{1} {1}_c\,\otimes\,\sigma_j\sigma_i\rho_t\sigma_i^\dagger\sigma_j^\dagger \nonumber\\
& & \quad + \sqrt{\gamma(1{-}\gamma)}\,\ketbra{0}{1}_c\,\otimes\,\sigma_i\sigma_j\rho_t\sigma_i^\dagger\sigma_j^\dagger\, \nonumber \\
& & \quad + \sqrt{\gamma(1{-}\gamma)}\,\ketbra{1}{0}_c\,\otimes\,\sigma_j\sigma_i\rho_t\sigma_j^\dagger\sigma_i^\dagger\,) \nonumber \\
& {=} & \ \ \, \sum_{i,j}\,p_iq_j\:\textswab{s:}[\sigma_i,\sigma_j](\rho_c\otimes \rho_t)
\label{Switch:qubit target}
\end{eqnarray}

\noindent From Eq. \ref{Switch:qubit target}, we notice that from pairwise combinations of $\sigma_i$ and $\sigma_j$ operators we can find out the output $\textswab{s:}[\mathcal{N}_{p},\mathcal{N}_{q}] \ (\rho_c\otimes \rho_t)$ and its minimum entropy $H^{\text{min}}(\textswab{s:}[\mathcal{N}_{p},\mathcal{N}_{q}])$, given we use an optimised target state $\rho_t$. Note that, the operations in $\{\sigma_i\}$ either commute or anti-commute. We can construct $\textswab{s:}[\sigma_i,\sigma_j](\rho_c{\otimes} \rho_t)$ by projecting the control qubit into the diagonal/antidiagonal basis which results to a product state. Denoting the anti-commutator by $\{$...$\}$ and the commutator by $[$...$]$, we have, with $\gamma{=}1/2$,  
$$
\textswab{s:}[\sigma_i,\sigma_j](\rho_c{\otimes} \rho_t) {=} \begin{cases} \ketbra{+}{+}_c{\otimes}\{\sigma_i,\sigma_j\}\rho_t\,\{\sigma_i,\sigma_j\}^\dagger  
\\ {\rm for}  [\sigma_i,\sigma_j]{=}0\\ \\
\ketbra{-}{-}_c{\otimes}[\sigma_i,\sigma_j]\rho_t\,[\sigma_i,\sigma_j]^\dagger \\ {\rm for}   \{\sigma_i,\sigma_j\}{=}0 \end{cases}.$$ The value of the Stokes parameter $S_2(\textswab{s:}[\sigma_i,\sigma_j])$, which is $1$ $(-1)$ for commuting (anti-commuting) operations in the ideal case, can be experimentally obtained by noting that $S_2(\textswab{s:}[\sigma_i,\sigma_j]) {=} \tr [( \sigma_1^c \otimes \sigma_0)\textswab{s:}[\sigma_i,\sigma_j](\rho_c{\otimes} \rho_t)]$, i.e., expectation value of the operator $\sigma_1^c\otimes \sigma_0$.

\subsection*{Experimental details}
\begin{figure}[!h]
\includegraphics[width=\columnwidth]{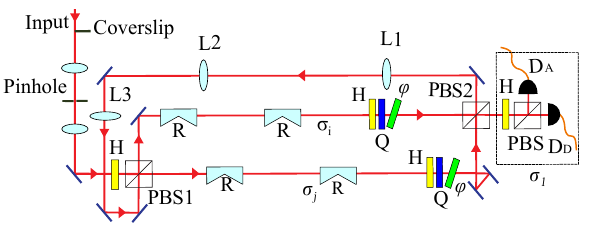}
\vspace{-3mm}
\caption{Schematic of quantum switch. R is the rotating prism \ref{eq:rot_prism}, H and Q are the half and quarter waveplates respectively and $\varphi$ are the phase plates. The control qubit is the polarisation, $\ket{\psi_c} {=} D$. The polarisation of the light controls order of Pauli operations $\{\sigma_{i}\}$ acting on the photonic spatial mode $\ket{\psi_t}{=}\mathrm{HG}_{10}$, for horizontal polarisation $H$, the order is $\sigma_{i}{\circ}\sigma_{j}$, for the vertical polarisation $V$ the order is $\sigma_{j}{\circ}\sigma_{i}$. The polarisation $D$ ensures superposition of the orders. $X$ is a polarisation measurement, determining the Stokes parameter of the measured photon in the diagonal/anti-diagonal basis. Lenses L1 to L3 form a telescope for mode-matching. \vspace{0mm}
}
\label{fig:Experimental_setup1}
\end{figure}

\noindent Our experimental setup is adapted from the quantum switch of Ref.~\cite{Goswami18}, removing cylindrical lenses and using only inverting prisms, since only Pauli operations where necessary, Fig.~\ref{fig:Experimental_setup1}. Our input light is diagonally-polarised ($| \rho_{c}\rangle {=} (|0\rangle {+} |1\rangle)/\sqrt{2}{=} D$), and is in the first-order  spatial mode ($| \rho_{t} \rangle {=} | 1  \rangle {=} \rm{| HG}_{10}  \rangle$).
 
\noindent We realise the unitary operations $\{\sigma_i\}$ using a pair of rotating prisms~\cite{Leach2004} as shown in Fig.~\ref{fig:Experimental_setup1}. A mechanical rotation of the inverting prism results in a rotation of the incoming spatial mode of the photon, the outputs of the combined operation $\{\sigma_i \circ \sigma_j \}$are shown in Fig.~\ref{fig:spatial_mode_and_target_qubit}.

To implement $\sigma_i, \sigma_j$, we use up to two rotating prisms~\cite{Goswami18} in each arm. Each prism is oriented at a physical angle $\theta$, which reflects and rotates an incoming beam by $2\theta$. The action of the rotating prism on our target qubit subspace is represented by the following unitary operation:
\begin{equation}
\textup{R}(\theta){=}\m{-\rm{cos} \,2\theta & \rm{sin}\,2\theta\\\rm{sin}\,2\theta & \rm{cos}\,2\theta}.
\label{eq:rot_prism}
\end{equation}
\noindent We impart the global phase $\varphi$ by a tilted phase plate. We write the transformation performed by the pair of prisms and the phase plate as:
\begin{equation}
U(\varphi,\theta_1,\theta_2){=}e^{i\varphi}\,\rm{R}(\theta_2)\,\rm{R}(\theta_1).
\label{eq:def_U}
\end{equation}
\noindent We place the phase plate only when we are doing $\sigma_2$ operation. For operations, $\sigma_1$ and $\sigma_3$ we need only one rotating prism, and for $\sigma_0$ and $\sigma_2$ we need both prisms. We achieve this by moving the second rotating prism via a translation stage. Table~\ref{table:implementation_unitaries} are the angles and phases used to implement the Pauli operations. Since this rotation also changes the polarisation of the field---the control qubit---which is not desired, this rotation is cancelled after the two rotating prisms by the actions of the half- (H), and quarter- (Q) waveplates and the phase plate ($\phi$), see Figure \ref{fig:Experimental_setup1}.

\begin{table}[t]
\centering
    \begin{tabular}{| c | c | c | c |}
    \hline
    Unitary & $\varphi$ & $\theta_1$ & $\theta_2$\\ \hline
    ${\sigma}_0$ & 0 & $\frac{\pi}{2}$ &$\frac{\pi}{2}$\\ 
    ${\sigma}_1$ & $0$ & $\frac{\pi}{4}$ & --\\
    ${\sigma}_2$ & $\frac{\pi}{2}$ & $\frac{\pi}{2}$ &$\frac{\pi}{4}$\\
    ${\sigma}_3$ & $0$ &$\frac{\pi}{2}$ & --\\
     \hline
   \end{tabular}
    \caption{Phases and angles for the unitary operations realised in our experiment, given by Eq.~\ref{eq:def_U}. \vspace{-2mm}}
    \label{table:implementation_unitaries}
\end{table}

We emphasise that in our architecture, the sender cannot access the individual channels---channels $\mathcal{N}_p$ or $\mathcal{N}_q$---without using the other one. That is, when we consider combinations of channels where one of the channels is unitary, this unitary channel cannot be accessed without also going through the noisy channel. This is certainly not a unique realisation of indefinite causal order. For example, one can use a Sagnac interferometer to achieve the same indefinite causal order as in Ref 12 of the main paper.

\subsection*{Effects of experimental imperfections}
\begin{figure}[!h]
\begin{center}
\includegraphics[width=\columnwidth]{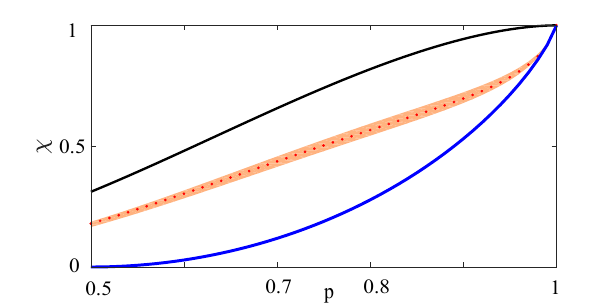}
\caption{Experimental Holevo capacity $\chi$ for $p{\geq}0.5$ region. where one channel is a bit-flip channel and the other one is a dephasing channel with varying strength $p$ as in the Eq. \ref{U_channel}. The solid black line is the theoretical predictions for Holevo capacity for indefinite causal order. The blue line is the theoretical prediction for Holevo capacity for definite causal order. The red dots are the experimental data points and the orange shade is the expected range due to non-ideal visibility. At $p {=} 0.5$, the experimentally measured Holevo capacity is $0.179{\pm}0.006$ bits, whereas $\chi_{d} {=} 0$.  }
\vspace{-6mm}
\label{fig:U_channel_expt}
\end{center}
\end{figure}

\noindent We note, the Pauli operations we are implementing by the rotating prisms, can be non-ideal due to uncertainty  of the angles.  This can affect the experiment in two ways. First, it leads to non-zero $S_1$ and $S_3$ Stokes parameters in the output control qubit and second, these nonzero terms contribute to  $H^{\text{min}}(\textswab{s:}[\mathcal{N}_{p},\mathcal{N}_{q}])$. To account for this issue, we note that the uncertainty associated with our rotation mounts is $\pm 1 \degree$.We numerically introduce random uniformly distributed error of $ \pm 1\degree$ to the angles of the rotating prisms. We repeat the simulation for 500 iterations and measure the capacity in each run. We confirm that the contribution of the random errors are well within the range of our experimental visibilities, which is reflected in the orange shades of the graphs in the Fig. \ref{depolarisation_graph}, \ref{1bit_graph}, \ref{1bit_graph_other_p}, and \ref{fig:U_channel_expt}. 

\subsection* {More general combinations of depolarising and dephasing channels}
\begin{figure}[!h]
\begin{center}
\includegraphics[width=\columnwidth]{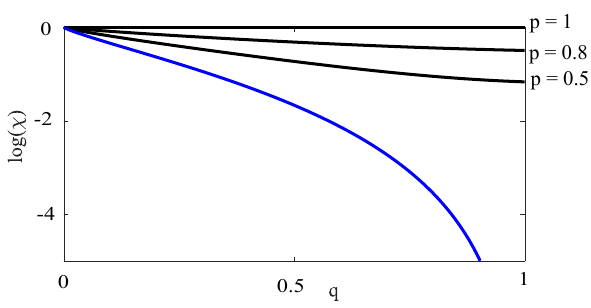}
\caption{Predicted logarithm of Holevo capacity $\chi$ for the channels in Eq. \ref{extreme_case_channels} vs. depolarisation noise strength, $q$. The solid black lines are for indefinite order in the cases of $p{=}0.5$, $p{=}0.8$, and $p{=}1$. The solid blue line are for definite order,  regardless of the value of $p$, the combination of the dephasing and depolarising channels scramble the information equally. At the full depolarisation strength, $q{=}1$, $\chi_{d}{=}0$ whereas $\chi_{i}$ are 0.31 bits, 0.61 bits, and 1 bit in case of $p{=}0.5$, $p{=}0.8$, and $p{=}1$ respectively. \vspace{-6mm}
}\label{fig:1bit_all_theory}
\end{center}
\end{figure}

\begin{figure}[!h]
\begin{center}
\includegraphics[width=\columnwidth]{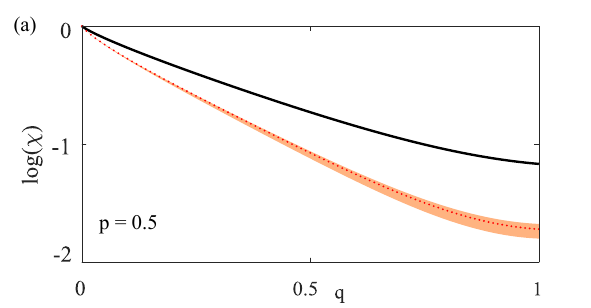}\\
\includegraphics[width=\columnwidth]{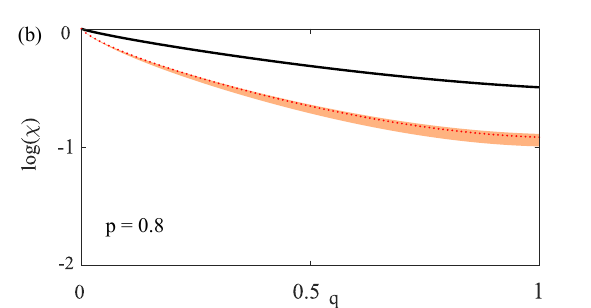}\\
\caption{ Logarithm of Holevo capacity $\chi$, where one channel is a depolarising channel of varying strength $q$, and the other one is a dephasing channel of strength $p$. The solid black lines are the theoretical predictions for Holevo capacity for indefinite causal order. The orange shades show the uncertainty due to non-ideal visibility in our experiment. The red dots are the experimentally measured data points. (a) shows the case when $p{=}0.5$. In this case the ideal value is $\chi_{i} {=} 0.31$ bits whereas we experimentally measure $0.179{\pm}0.006$ bits. (b) is the case when $p{=}0.8$. Here, the $\chi_{i}$ is 0.61 bits and we measure $0.42{\pm}0.01$ bits. }
\vspace{-3mm}
\label{1bit_graph_other_p}
\end{center}
\end{figure}

\noindent In the main section we have introduced a combination of depolarising channel of strength $q$ and a dephasing channel of strength $p$ as shown in Eq.\ref{extreme_case_channels}. We have shown the special case for $p {=} 1$ where it is possible to achieve unit Holevo capacity. In this section we show more general situations where $p \leq 1$. In Fig. \ref{fig:1bit_all_theory}, the black lines show the cases for either $p {=} 0.5$, $p {=} 0.8$, or $p {=} 1$ combined with a depolarising channel of strength $q$ in an indefinite order. The blue line shows the capacity when these channels are in definite order, which regardless of the value of $p$ monotonically decreases to zero when $q{=}1$. In contrast, with indefinite order arrangements the predicted Holevo capacities are $0.31$ bits, $0.61$ bits and $1$ bit for $p{=}0.5$, $p{=}0.8$ and $p{=}1$ respectively. In Fig. \ref{1bit_graph_other_p} we plot the logarithm of Holevo capacities as a function of both depolarising channel strength $q$ for 2 different strengths of the dephasing channel, $p{=}0.5$, and $p{=}0.8$. The black curves on the graphs show the predicted Holevo capacity for the indefinite order, the orange shade is the uncertainty due to the fringe visibility and the red dots are experimental data points. In the case of $p{=}0$, the dephasing channel behaves as an identity channel and the input state $\rho_t$ is equally scrambled in both definite and indefinite order cases, thus resulting no overall advantage. However, in other cases there is non-zero information-theoretic advantage. 
At a full depolarising strength of $q{=}1$: in the case of $p{=}0.5$, the predicted  $\chi_{i}{=}0.31$  bits, whereas we experimentally show $0.179{\pm}0.006$ bits; at $p{=}0.8$, the predicted $\chi_{i}{=}0.61$ bits, we experimentally show $0.42{\pm}0.01$ bits.

\vspace{2mm}

\subsection* {Bit flip and bit-phase-flip channels}

Consider two noisy channels, a bit-flip and a bit-phase-flip channel with strength $p$ given, correspondingly, by 
\begin{equation}
   \begin{array}{l}
N_{p}^{1}(\rho_t){=}(1{-}p)\rho_t{+}p\sigma_{1}\rho_t\sigma_{1},\\
N_{p}^{2}(\rho_t){=}(1{-}p)\rho_t{+}p\sigma_{2}\rho_t\sigma_{2}.\\
  \end{array}
  \label{U_channel}
 \end{equation}
Then,
\begin{equation}
  \begin{array}{l}
\epsilon_{+}(\rho_t)=(1{-}p)^2\rho_t {+} p(1{-}p)(\sigma_1\rho_t\sigma_1{+}\sigma_2\rho_t\sigma_2)\\
\epsilon_{-}(\rho_t){=}p^2\sigma_3\rho_t\sigma_3.
\end{array}
\end{equation} 

\begin{figure}[!t]
\begin{center}
\includegraphics[width=\columnwidth]{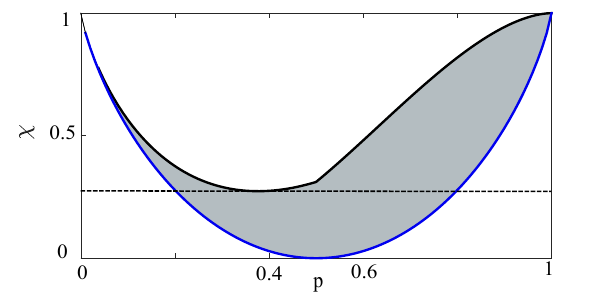}\\
\caption{Predicted Holevo capacity $\chi$ for the channels in Eq. \ref{U_channel} versus noise parameter, $p$. The solid black line is the Holevo capacity of depolarising channels in indefinite order and the solid blue line is for the case of definite order. We observe a knee at $p{=}0.5$, this is because the optimum state for $p{\leq}0.5$ are the eigenvectors of the operator $\sigma_{1}$ and $\sigma_{2}$. On the other hand, for $p{\geq}0.5$ the optimum state is the eigenvector of $\sigma_3$. At $p{=}0.5$, the definite-ordered channel becomes completely depolarising, making $\chi_{d}=0$, whereas at this point $\chi_{i}=0.31$ bits. In the region of $p{\geq}0.5$, the maximum difference between the two cases is 0.55 bits, occurring when $p$=0.75.  The  black horizontal dashed line denotes the minimum capacity of the indefinite-ordered case, 0.27 bits when $p$=0.37.
\vspace{-6mm}}
\label{fig:U_channel}
\end{center}
\end{figure}

\noindent It should be highlighted that the outcome corresponding to the $\ket{-}\bra{-}$, namely the $\epsilon_{-}(\rho_t){/}p^2$ channel, is actually a purely unitary channel, capable of achieving perfect transmission despite both channels being noisy  (as was also pointed out recently in Ref.\cite{Salek2018}). However, here we are interested in the Holevo capacity resulting from considering both possible outcomes weighted by their corresponding probabilities, rather than the post-selected $\epsilon_{-}(\rho_t)$. We solve the corresponding minimization problem to compute the Holevo capacity.  Fig. \ref{fig:U_channel}  plots the Holevo capacity $\chi$ against different noise parameters $p$ for both the cases of noisy channels in a definite order, and in an indefinite order. The black curve is the Holevo capacity for indefinite order, while the blue curve is for definite order. As shown by the difference of these two curves, indefinite order provides an advantage over the definite-order case. Indefinite order allows transmission of $0.27$ bits or more over the whole range of $p$ values.  The minimum capacity of $0.27$ bits which occurs at $0.37$, is higher than the capacity of the definite-order case from $p{=}0.2$ to $p{=}0.8$, $60\%$ of the range of $p$. The maximum advantage $\rm max[(\chi_{i} {-} \chi_{d})] {=} 0.55$ bits occurs at $p {=} 0.75$, where the Holevo capacity for the quantum switch and definite order cases are $0.75$ bits and $0.19$ bits, respectively. 
\noindent We can see there is a knee in the $\chi_{i}$ at noise parameter $p {=} 0.5$. This is because when $p{\leq}0.5$, the eigenvectors of $\sigma_1$ or $\sigma_2$ achieve the minimum entropy $H^{\text{min}}(\textswab{s:}[\mathcal{N}_{p}^{1},\mathcal{N}_{p}^{2}])$, while for $p{\geq}0.5$, the corresponding optimum state turns out to be the eigenvectors of $\sigma_3$. When $p{=}0.5$, where the classical order results in full depolarization, the capacity resulting from indefinite order is $\chi_{i}{\approx} 0.31$ bits. In this case, any of the eigenstates of the Pauli operator achieves the maximum capacity thus increasing the domain over which we can optimally encode the input.

\subsection*{Holevo capacity from path superposition}
\noindent Although the advantage in communication is not unique to superposition of order, there is strong reason to believe that the perfect transmission discussed above is not possible via a superposition of paths, where we place two noisy channels in two arms of an interferometer \cite{Branciard2018}. The optimised Holevo capacity that we obtain for a fully depolarising channel and a unitary channel in path superposition is 0.75 bits. Moreover, we prove that reaching a value of 1 bit is impossible: 

\noindent In our communication task, we consider the target system, the control system and two Pauli channels to transfer the target qubit. The control enables the target to go through a superposition of paths.  In our experiment, $|\psi_c\rangle$ is set to be $|+\rangle$, the target state is optimized to achieve the maximum communication capacity, and a generalized measurement on the control system is applied after noisy channels.

\noindent We consider our Pauli channels to be the depolarising channel, $\mathcal{N}$, as shown in Eq. \ref{depolarising channels}, and $\sigma_3$.  Following \cite{Branciard2018}, after tracing out the environment, the output control and target state takes the following form: 
\begin{align} 
\rho_{tot}^{(ct)} &{=} \frac{1}{2} \Big( |0\rangle\langle 0|^c\otimes \mathcal{N}(\rho_{t})+|1\rangle\langle1|^c\otimes \sigma_3\rho_{t}\sigma_3 \\ \nonumber
&+|0\rangle\langle1|^c\otimes T_0\rho_{t}\sigma_3 +|1\rangle\langle0|^c\otimes \sigma_3 \rho_{t}T_0^{\dagger} \Big)
\end{align}
where $T_0{=}\sum_i { e_i  K_i}$,  $K_i$ are arbitrary choices for Kraus operators for the depolarising channel with the constraint $ \sum_i |e_i|^2 {=}1$. Exploiting the unitary freedom in the operator-sum representation (see for example Ref.~\cite{chuang00}), one has that 
$e_i K_i {=}e_i \sum_j U_{i,j} \frac{\sigma_j}{2}$, with $U_{i,j}$ the entries of a unitary matrix, and thus  
$T_0 {=} \sum_{j} (\sum_i e_i U_{i,j})\frac{\sigma_j }{2} \equiv \sum_j f_j \sigma_j/2$, with $\sum_j |f_j|^2 =1$. 
 Now if we measure the control qubit in the $\{M_+,M_-\}$ basis , 
\begin{equation}
\begin{array}{ll}
M_+{=}\cos\alpha|0\rangle+\sin\alpha|1\rangle,\\
M_-{=}-\sin\alpha|0\rangle+\cos\alpha|1\rangle,
 \end{array}
\end{equation}

the two possible output states, with corresponding probabilities $p_+$ and $p_-$, are given by 

\begin{align}
p_+ \rho^{t}_+&=\frac14[\cos^2\alpha\frac{\sigma_0}{2}+\sin^2\alpha\sigma_3\rho_{t}\sigma_3 \nonumber \\ \nonumber
 \, \, \, &+\frac{\cos{\alpha}\sin{\alpha}}{2}\sum_i(f_i \sigma_i\rho_{t}\sigma_3+f_i^{*} \sigma_3\rho_{t}\sigma_i)],\\
p_- \rho^{t}_-&{=}\frac14[\sin^2\alpha\frac{\sigma_0}{2}+\cos^2\alpha\sigma_3\rho_{t}\sigma_3 \\ \nonumber &-\frac{\cos\alpha\sin\alpha}{2}\sum_i(f_i \sigma_i\rho_{t}\sigma_3+f_i^{*} \sigma_3\rho_{t}\sigma_{i})].
 \end{align}
where $f_i^{*}$ is the complex conjugate of $f_i$.

Now, achieving a unit Holevo capacity implies that both the input and output state must be pure. In the Bloch representation, this implies that 

\begin{align}
 \sum_i |r_i^{(+)}|^2{=}1, \label{outp+} \\
 \sum_i |r_i^{(-)}|^2{=}1, \label{outp-} \\
 \sum_i |r^{(in)}_i|^2{=}1 \label{outin}
\end{align}

\noindent where $\{r_i^{(+)}\}$, $\{r_i^{(-)}\}$ and $\{r_i^{(in)}\}$ are the Bloch vector components of the projected states $\rho^t_{+}$, $\rho^t_{-}$ and the input target state, $\rho_t$ respectively.  We proceed by writing $f_i$ into its real and imaginary parts, i.e., $f_j{=} f_{j,R} + i f_{j,I}$. Then, from $\sum |r_i^{(+)}|^2 {=}\sum |r_i^{(-)}|^2$ we find that
\begin{align}
\sin{2\alpha} \,  (r_3^{(in)} f_{0,R}+f_{3,R}+r_1^{(in)} f_{2,I}-r_2^{(in)} f_{1,I})+\cos{2\alpha}{=}0
\end{align}
which can be replaced into $\sum |r_i^{(+)}|^2 +\sum |r_i^{(-)}|^2 {=}2$ to obtain
\begin{widetext}
\begin{align}
\nonumber &\,\, \sin^2{ 2\alpha}\,  [(r_2^{(in)} f_{0,I}-r_3^{(in)} f_{1,R}+r_1^{(in)} f_{3,R}+f_{2,I})^2 +(r_1^{(in)} f_{0,I}+r_3^{(in)} f_{2,R}+f_{1,I}-r_2^{(in)} f_{3,R})^2 \\ & +(f_{0,R}+r_1^{(in)} f_{1,R}+r_2^{(in)} f_{2,R}+r_3^{(in)} f_{3,R})^2]+\cos{ 4\alpha} {=} 2
\label{eq1}
\end{align}
\end{widetext}
Next we proceed to show that,  Eq.~\ref{eq1} cannot hold under conditions ~\ref{outin} and $\sum_i |f_i|^2 {=} \sum_i\sum_{k\in\{R,I\}} {f_{i,k}}^2{=}1$. Note that for this purpose, it is sufficient to prove that 
 \begin{equation}
   \begin{array}{ll}
     F(r,f)&=(r_2^{(in)} f_{0,I}-r_3^{(in)} f_{1,R}+r_1^{(in)} f_{3,R}+f_{2,I})^2\\
     &+(r_1^{(in)} f_{0,I}+r_3^{(in)} f_{2,R}+f_{1,I}-r_2^{(in)} f_{3,R})^2 \\ 
     & +(f_{0,R}+r_1^{(in)} f_{1,R}+r_2^{(in)} f_{2,R}+r_3^{(in)} f_{3,R})^2<3.
  \end{array}
\end{equation}

Exploiting the relation that $2xy\leq x^2+y^2$ for any real $x$ and $y$, Eq.~\ref{outin}, and that for 
that for $|r_i|\leq 1$ one has that $r_i \leq |r_i|$ and $ |r_i|^2 \leq |r_1|$, 
we have   \begin{widetext}
  \begin{equation}
   \begin{array}{ll}
     F(r,f)&\! \leq (1+r_1^{(in)}+r_2^{(in)}+|r_3^{(in)}|^2)f_{0,I}^2+(1+r_1^{(in)}+r_2^{(in)}+r_3^{(in)})(f_{0,R}^2+f_{1,I}^2+f_{2,I}^2+f_{3,R}^2)\\
     &+(1+r_1^{(in)}+|r_2^{(in)}|^2+r_3^{(in)})f_{1,R}^2+(1+|r_1^{(in)}|^2+r_2^{(in)}+r_3^{(in)})f_{2,R}^2\\
     &\leq(1+|r_1^{(in)}|+|r_2^{(in)}|+|r_3^{(in)}|)\sum_i|f_i|^2\\
     &\leq 1+\sqrt{3 \sum_{i} |r_{i}^{(in)}|^2} \\
     & \leq 1 +\sqrt{3}<3,
     \end{array}
   \end{equation}
 \end{widetext}
as desired. This supports the idea that $\mathcal{N}$ and a unitary channel in a superposition of paths cannot lead to a pure output and thus cannot achieve unit Holevo capacity.

\end{document}